\documentclass[manuscript]{aastex}
\slugcomment{}

\shorttitle{Extrasolar Planet Angular Momentum}
\shortauthors{Armstrong et al.}

\begin{document}

%% LaTeX will automatically break titles if they run longer than
%% one line. However, you may use \\ to force a line break if
%% you desire.

\title{Specific Angular Momentum of Extrasolar Planetary Systems}

%% Use \author, \affil, and the \and command to format
%% author and affiliation information.
%% Note that \email has replaced the old \authoremail command
%% from AASTeX v4.0. You can use \email to mark an email address
%% anywhere in the paper, not just in the front matter.
%% As in the title, use \\ to force line breaks.

\author{John C. Armstrong}
\affil{Department of Physics, Weber State University, Ogden, UT, 84408-2508, \\ jcarmstrong@weber.edu}

\author{Shane L. Larson}
\affil{Department of Physics, Utah State University}

\author{Rhett R. Zollinger}
\affil{Department of Physics and Astronomy, University of Utah}

%% Notice that each of these authors has alternate affiliations, which
%% are identified by the \altaffilmark after each name.  Specify alternate
%% affiliation information with \altaffiltext, with one command per each
%% affiliation.

%% Mark off your abstract in the ``abstract'' environment. In the manuscript
%% style, abstract will output a Received/Accepted line after the
%% title and affiliation information. No date will appear since the author
%% does not have this information. The dates will be filled in by the
%% editorial office after submission.

\begin{abstract}

As the number of known planetary systems increases, the ability to follow-up and characterize the extent of any system becomes limited.  This paper considers the use of specific angular momentum as a metric to prioritize future observations.  We analyze 431 planets in 367 known extrasolar planetary systems from  \cite{2006ApJ...646..505B}  (including updates to their online catalog, current to April, 2011) and estimate each system's orbital angular momentum.  The range of partitioning of specific angular momentum in these systems is found to be large, spanning several orders of magnitude.  The analysis shows that multi-planet systems tend to have the highest values of specific angular momentum normalized against the planetary masses.  This suggests that in high angular momentum systems, the dominant contributors have already been discovered, and that single-planet systems with low observed angular momentum may be the most likely candidates for additional undiscovered companions compared to their high angular momentum, single-planet counterparts.  The multi-planet system, GJ 581, is considered as a historical case study to demonstrate the concept, examining how the specific angular momentum of the know planetary system evolved with each discovery.

\end{abstract}

%% Keywords should appear after the \end{abstract} command. The uncommented
%% example has been keyed in ApJ style. See the instructions to authors
%% for the journal to which you are submitting your paper to determine
%% what keyword punctuation is appropriate.


%% From the front matter, we move on to the body of the paper.
%% In the first two sections, notice the use of the natbib \citep
%% and \citet commands to identify citations.  The citations are
%% tied to the reference list via symbolic KEYs. The KEY corresponds
%% to the KEY in the \bibitem in the reference list below. We have
%% chosen the first three characters of the first author's name plus
%% the last two numeral of the year of publication as our KEY for
%% each reference.

%% Authors who wish to have the most important objects in their paper
%% linked in the electronic edition to a data center may do so by tagging
%% their objects with \objectname{} or \object{}.  Each macro takes the
%% object name as its required argument. The optional, square-bracket 
%% argument should be used in cases where the data center identification
%% differs from what is to be printed in the paper.  The text appearing 
%% in curly braces is what will appear in print in the published paper. 
%% If the object name is recognized by the data centers, it will be linked
%% in the electronic edition to the object data available at the data centers  
%%
%% Note that for sources with brackets in their names, e.g. [WEG2004] 14h-090,
%% the brackets must be escaped with backslashes when used in the first
%% square-bracket argument, for instance, \object[\[WEG2004\] 14h-090]{90}).
%%  Otherwise, LaTeX will issue an error. 

\section{Introduction}

In 2006, \citeauthor{2006ApJ...646..505B} published a catalog of 172 known low-mass companions.   The web version of this catalog ({\em http://exoplanets.org/planets.shtml}) has been updated to include 431 planets in 367 systems as of April 18, 2011.  With such an extensive catalog, it is possible to perform a meta-analysis of these objects to uncover some of the general properties of solar system formation.  Previous studies have already analyzed the distribution of masses, eccentricities, orbital distances \citep{2006ApJ...646..505B, 2005A&A...431.1129H} and stellar properties \citep{2007MNRAS.378.1141G, 2004A&A...417..353E, 2003A&A...398..363S}.  However, a comprehensive study of the distribution of each system's angular momentum has yet to be performed.

By studying the angular momentum to other extrasolar planets, we can examine the nature of these systems and probe their formation mechanisms.  Similar work has been performed on the combined orbital properties of the extrasolar planets.  For example, \citet{2003A&A...407..369U} examined the distribution of orbital periods and masses of the extrasolar planets as they relate to planet migration theory.  They discovered that the so-called ``Hot Jupiters'' tended to be lower mass planets  compared to planets more distant from their central stars (see Figure 1. in \citeauthor{2003A&A...407..369U}), and also noticed several gaps in the period distribution indicating that planet migration is mass and distance dependent.  The goal of this study is to explore these ideas in more detail, by comparing the specific angular momentum of planets across systems, thus removing the mass dependence.

Casting the problem in the light of a two-body central potential for those systems with only one known planet, we can estimate the orbital angular momentum for each system.  Since most of the planets have been discovered using the radial velocity technique \citep{2006ApJ...646..505B}, we have the minimum mass, $M_{p} sin(i)$, the  orbital properties (semimajor axis, eccentricity, argument of perihelion, and orbital period) as well as the properties of the star (star type and mass, and an estimate of the star's radius).  In \S 2, we describe the catalog and our selection of systems to use for this study.  We also discuss both the analytical, two body orbital angular momentum calculation, and a numerical calculation to extend the analysis to multi-planet systems.  In \S 3 we show the distribution of angular momentum with respect to planet and stellar mass and examine a case study in \S4. The implications of this work are outlined in \S 5.

\section{Angular momentum calculations}

For this study, we took a sample of 367 planetary systems, with 431 planets from \cite{2006ApJ...646..505B}, including web updates to this catalog current as of April, 2011.  We excluded one system, $HD 17092b$, from the study because it lacked an estimate for the mass of the parent star.  For our calculations, we required the stellar mass and the planet mass for each member of the system, along with their eccentricity, orbital period, semimajor axis, and argument of perihelion.  The angular momentum of single planet systems were first computed analytically for verification.  Next, all of the systems were numerically integrated to compute the total orbital angular momentum.

\subsection{Analytical calculation of orbital angular momentum}

Treating each planet-star system as a two-body central force problem we calculated the system's total orbital angular momentum, 

\begin{equation}
L = \mu \sqrt{GMa(1-e^2)},
\end{equation}

\noindent where $\mu$ is the reduced mass, $G$ is the universal gravitational constant, $M$ is the combined stellar and planetary mass, $a$ is the semi-major axis and $e$ is the eccentricity. This provides the total orbital angular momentum only, excluding any rotational angular momentum of the parent star.  Since we only know the minimum mass in most cases, however, this can be approximated by

\begin{equation}
L_{obs} \cong L sin(i),
\end{equation}

\noindent for a given inclination, $i$, assuming the mass of the planet is much less than the mass of the parent star.  The specific orbital angular momentum, $J$,  of the system is 

\begin{equation}
J = L_{obs}/M,
\end{equation}

\noindent where $M$ can be replaced by either the mass of the planet, $M_P$, or the mass of the star, $M_*$, depending on the analysis.  This analytical method provides a check against the accuracy of the numerical simulations, and illustrates the dependence of angular momentum on planet mass ($L \propto M_P$) and distance from the star ($L \propto \sqrt{a}$), as well as the dependence of specific angular momentum on stellar mass ($J_P \propto M_*^{1/2}$ if normalized by the planet mass, and $J_* \propto M_*^{-1/2}$ if normalized by the stellar mass).

\subsection{Numerical calculations}

To include multi-planet systems with the two-body systems, we numerically integrated all systems over 1000 years using the symplectic integrator {\em Mercury} \citep{1999MNRAS.304..793C}.  Each of the calculations used the measured values of semimajor axis, eccentricity, and argument of perihelion for each planet in the system.  In the case where either the eccentricity or the argument of perihelion were unknown, they were assumed to be zero.  In addition, the mutual inclination of multi-planet systems was set to zero, and all planets started the integration at perihelion.  The minimum timestep for each integration was 0.02 days (roughly $1/200^{th}$ of the shortest orbital period in the sample.  The output was recorded every 100 days during the simulation.  

During the integration, each system was was monitored to insure both energy and momentum were conserved, and the last computed heliocentric position and velocity of each planet were used to compute the total angular momentum of the system.  The total angular momentum was use to derive two specific angular momenta: $J_P$, the total orbital angular momentum divided by the sum of the planetary masses in the system, and $J_*$, the total angular momentum divided by the mass of the parent star.  The numerical results for two-body systems were validated against the analytical calculations outlined above.

%%The results of these simulations are summarized in the Table \ref{tab:results}, with both $J_P$ and $J_*$ for each system.  

Figure 1 shows the numerical results for $J_P$ as a function of stellar mass in four cases: for all systems, for multi-planet systems, for systems with total planet mass more than $2.0 \ M_J$, and for systems with total planet mass less than $2.0 \ M_J$.  Figure 2 is the same for $J_*$.  For comparison in both figures, the solid line represents the specific angular momentum of a single Jupiter mass planet on a circular orbit at $5.2 \ AU$ around the parent star.  Figure 3 plots $J_P$ vs. $J_*$ for the same four cases.

\section{Results}

Figures 1-3 detail the distribution of angular momentum in these systems, both normalized against planet mass and stellar mass.  In Figure 1, the  majority of systems have $J_P$ less than the somewhat arbitrary Jupiter model limit.  Also, systems with low total planet mass tend have lower angular momenta, and systems with higher masses have higher angular momenta, independent of stellar distance.  

The specific angular momenta in Figure 2 tells a parallel story.  Again, we see that 67\% of the multi-planet systems lie above the Jupiter model line.  In addition, the systems with higher total planetary mass are also above the Jupiter model line, and those systems with lower total planetary mass are below it when the specific angular momenta are normalized against the stellar mass.  Figure 3 shows the trend between $J_P$ and $J_*$ for each system.  As expected, we see similar relationships. 

Two types of systems are found on or above the Jupiter model line: systems with a single large planetary mass, and those systems with multiple planets. The Jupiter comparison, while arbitrary, seems significant.  It is certain that Jupiter is the highest mass planet in our system, and dominates the orbital angular momentum.  The calculations indicate that multi-planet systems tend to have the highest values of $J_P$.  Since we are fairly certain we have found the highest mass planet in the multi-planet systems, this indicates a potential upper limit on the total specific angular momenta for planetary systems, when normalized against the planet mass.  Additionally, in systems with one large planetary mass, we have likely already discovered the dominant contribution to the angular momentum in the system.  This indicates that single-planet systems with low observed angular momentum may be likely candidates for additional undiscovered companions compared to their high angular momentum single-planet counterparts.

The full range of the specific angular momenta of the systems spans several orders of magnitude, with the high planetary mass systems above the Jupiter model line, and low mass systems below it.  To further explore this, we generated histograms of the known planetary properties for those systems with angular momenta above $10^{13} \ m^{2} \ s^{-1}$ (corresponding to the specific angular momentum of the Jupiter-Sun system).  Figure 5 shows the fraction of systems for the planet mass (top panel) and the stellar mass (bottom panel).  The solid line is the total sample, the dashed line the system with specific angular momentum less than $10^{13} \ m^2 \ s^{-1}$, and the dotted line the systems with specific angular momentum greater than $10^{13} \ m^2 \ s^{-1}$.

Again, according to Figure 5, the lower mass planets in the sample (those below $2.0 \ M_J$) tend to have low values of the specific angular momenta.  However, there appears to be no pattern with respect to stellar mass.

\section{A Case Study}

\begin{table}[h]
\caption{\narrower A case study of GJ 581 for the four known planets as well as the two unconfirmed planets. The table lists the $\log\left(J_P\right)$ as well as $J_P$ relative to the specific angular momentum of a system with only a Jupiter-like planet orbiting at 5.2 AU, denoted as $J_{P,Jupiter}$} 
\label{tab:case_study}
\centering
\begin{tabular}{lcc} \hline
 {\em System }     &  {\em $\log\left(J_P\right)$} & {\em $\frac{J_P}{J_{P,Juipter}}$} \\ \hline \hline
GJ 581 b & 14.70 & 0.09 \\                         
GJ 581 b,c & 15.06 & 0.20 \\                         
GJ 581 b-d & 15.35 & 0.39 \\                         
GJ 581 b-e & 15.42 & 0.47 \\                         
GJ 581 b-f & 15.67 & 0.85 \\                         
GJ 581 b-g & 15.75 & 1.00 \\   
\hline
GJ 581 with "Jupiter" & 15.75 & 1.00 \\                      
\hline \hline
\end{tabular}
\end{table}

As noted above, single-planet systems with relatively low angular momentum may have additional planets yet to be discovered compared to their high angular momentum counterparts.  To illustrate this point, we took a closer look at the multi-planet system, GJ 581.  As of 2009, this system has four confirmed planets, GJ 581 b, c, d and e.  Also, \citet{2010ApJ...723..954V} claim the discovery of two additional unconfirmed planets g and f.

To analyze this system in light of its history of discovery, we ran additional N-body simulations of the system in the following configurations: GJ 581 b alone; b and c; b through d; b through e; b through f; and one simulation with all six planets. Table \ref{tab:case_study} lists the $\log\left(J_P\right)$ as well as $J_P$ relative to the specific angular momentum of a system with only a Jupiter-like planet orbiting at 5.2 AU, denoted as $J_{P,Jupiter}$.  Note that as additional planets are discovered, the value of  $\log\left(J_P\right)$ approaches the Jupiter model line indicated in Figure 1 for a star with $\log\frac{M_*}{M_{sun}} = -0.5$.

This case study reveals that, as new planets are discovered, the total planet-mass weighted angular momentum, $J_P$, increases. In 2005, GJ 581 b gave the system only  9\% of what one would expect from a Jupiter analog system.  By 2007, when c and d were discovered, that value reached 39\%.  By 2009, with the discovery of e, the value was as high as 47\%.  By this analysis, the unconfirmed planets ``complete the system''.  In fact, even if these planets are refuted, GJ 581 seems a likely candidate for additional planetary discovery.

\section{Discussion}

The analysis outlined above leads us to the following observations:

\begin{itemize}
	\item Based on the extrasolar planet catalog and an analysis of our solar system, there appears to be a limit to the amount of angular momentum that can be expected in an extrasolar planet system.  Multi-planet systems, or systems with very massive planets, approach this limit which is very close to the limit defined by the specific angular momentum of Jupiter in our solar system.
	\item Planets with masses less than $2.0 \ M_J$ have the lowest total star-normalized specific angular momentum, independent of semimajor axis.
	\item Planets with masses greater than $2.0 \ M_J$ have high total star-normalized specific angular momentum, independent of semimajor axis.
	\item Most multi-planet systems lie on or above the Jupiter model line in both $J_P$ and $J_*$ specific angular momenta.  In the case of $J_P$, the standard deviation around the Jupiter model line for the multi-planet systems is $\sigma = 3.9 \times 10^{15} \ m^2 \ s^{-1}$.  55 Cnc, with five planets, lies $3\sigma$ from that line, but $\tau$ Boo, with relatively low $J_P$, is over $12\sigma$ away from the Jupiter model line.
	\item High angular momentum systems may fall into two categories: systems where one large planetary mass dominates, or systems with multiple planets.  This potentially indicates that systems with low angular momenta provide the best targets for future study, as they may have a higher potential to contain more undiscovered companions.
	\item There is little dependence of specific angular momentum on stellar mass.
	\item Single planet systems with low measured planet-mass weighted specific angular momentum may be likely targets for further planets to be discovered.
\end{itemize}

These observations, combined with previous work on the mass dependence of planetary migration \citep{2003A&A...407..369U}, show the low mass ``Hot Jupiters'' lost large amounts of angular momentum compared to their high mass counterparts, which do not undergo migration.  However, this also shows that the mass dependence on the migration is independent of other orbital parameters, as the angular momentum calculation specifically includes the eccentricity.  Highly eccentric orbits, in principle, can also lower angular momentum, although mass is by far a more important factor (it takes an eccentricity of 0.7 to reduce the angular momentum by a factor of 2).  Still, including the orbital properties, the mass cutoff for this migration appears to be $2 \ M_J$, with low mass systems losing $10 \ m^{2} \ s^{-1}$ to $1000 \ m^{2} \ s^{-1}$ of specific angular momentum during migration.  In addition, \citet{2004ApJ...604..388I} have linked planetary migration rates to planet mass, and have numerically estimated the distribution of planet masses as a function of semimajor axis.  The angular momentum distribution further illustrates this point.  Taking the time derivative of Eq. (1), and using the fact that the migration time is inversely proportional to the rate of change in semimajor axis, $\dot{a}$ \citep{2004ApJ...604..388I}, leads to 

\begin{equation}
\frac{\partial L}{\partial t} \propto \dot{a} \propto \frac{1}{\tau_{mig}}
\end{equation}

\noindent Since the migration time is proportional to the planet mass, larger mass planets have lost less angular momentum during the same period of time, and migrated less than their low mass counterparts.  The implication here is that, since the angular momentum flux is low, high mass planets on small orbits must have formed {\em in situ} and not moved there via migration.  Since the close-in planets are all low mass, migration is implicated in their formation process.

Observational biases exist in the extrasolar planet database.  The inclination of the system, unknown in the vast majority of cases, will tend to increase our calculated values of the the total angular momentum.  However, systems inclined 45 degrees would only increase the angular momentum by 42 \%, which is small considering the angular momentum ranges over three orders of magnitude.  

In addition, after 12 years of searching, observers are just now probing Jupiter-mass companions at Jupiter-like distances.  At the moment, the angular momentum calculations presented here are dominated by the planet mass.  Systems with low mass, but a larger semimajor axis, would also produce high angular momentum systems by a roughly proportional amount give the ranges of masses and semimajor axes in question.  The dearth of low-mass, high angular momentum systems in our study may represent this bias.  However, the ``Hot Jupiters'' of the extrasolar planet parameter space are the most well sampled and search techniques are most sensitive to high-mass planets.  If there were close-in, high-mass planets that were subject to angular momentum loss through migration, we would see them in this sample.

There are a number of additional questions to be addressed following this initial study.  As it stands, existing, undiscovered planets in these 190 systems might account for some of the``missing" angular momentum.  We suggest that the results from this paper might be used to prioritize future observations in an effort to fill in the missing pieces of the story.   

\section*{Acknowledgments}

We thank Weber State University's Ott Planetarium and acknowledge NASA grant NNX06AE61G, {\em Project PLANET: Planetarium Learning and New Education Technology}, for funding WSU's 132-processor computing cluster used in this study, as well as the collaborative environment provided by WSU's Scientific Analysis and Visualization Initiative.  We are grateful to Brad Carroll and Dale Ostlie for Problem 2.6 in {\em Modern Astrophysics} \citep{2006ima..book.....C} which was the motivation behind this work and the original inspiration for Figure 4, which was based on Figure 23.8 on page 858 of their text.  We would also like to thank the 2006 Astrophysics class who willingly extended this problem to cover the known extrasolar planets (for extra credit, of course).  
%% To help institutions obtain information on the effectiveness of their
%% telescopes, the AAS Journals has created a group of keywords for telescope
%% facilities. A common set of keywords will make these types of searches
%% significantly easier and more accurate. In addition, they will also be
%% useful in linking papers together which utilize the same telescopes
%% within the framework of the National Virtual Observatory.
%% See the AASTeX Web site at http://www.journals.uchicago.edu/AAS/AASTeX
%% for information on obtaining the facility keywords.

%% After the acknowledgments section, use the following syntax and the
%% \facility{} macro to list the keywords of facilities used in the research
%% for the paper.  Each keyword will be checked against the master list during
%% copy editing.  Individual instruments or configurations can be provided 
%% in parentheses, after the keyword, but they will not be verified.

% {\it Facilities:} \facility{Nickel}, \facility{HST (STIS)}, \facility{CXO (ASIS)}.

%% Appendix material should be preceded with a single \appendix command.
%% There should be a \section command for each appendix. Mark appendix
%% subsections with the same markup you use in the main body of the paper.

%% Each Appendix (indicated with \section) will be lettered A, B, C, etc.
%% The equation counter will reset when it encounters the \appendix
%% command and will number appendix equations (A1), (A2), etc.
%\bibliographystyle{icarus} 
%\bibliography{/Users/jca/Projects/Papers/masterbiblio}

\clearpage

\begin{figure*}
\epsscale{1.0}
\label{fig:specplanet}
\plotone{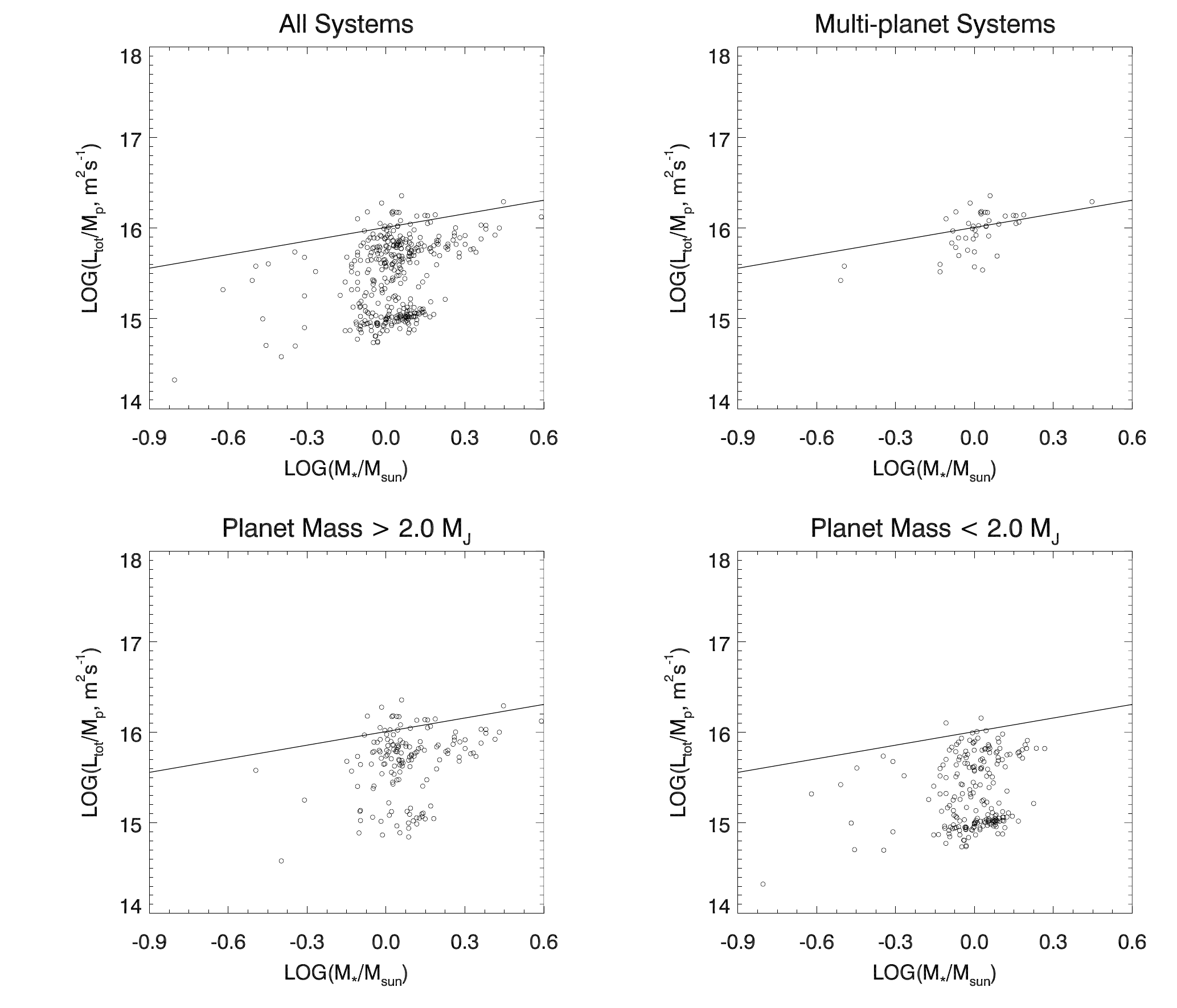}
\caption{Plot of the specific angular momentum, normalized to the total planetary mass of each system, for all systems, multi-planet systems, systems with total planetary mass above 2.0 $M_J$, and systems with total planetary mass below 2.0 $M_J$.   The solid line represents a model of a Jupiter mass planet in a circular orbit at 5.2 $AU$ for comparison.}
\end{figure*}

\begin{figure*}
\epsscale{1.0}
\label{fig:specstar}
\plotone{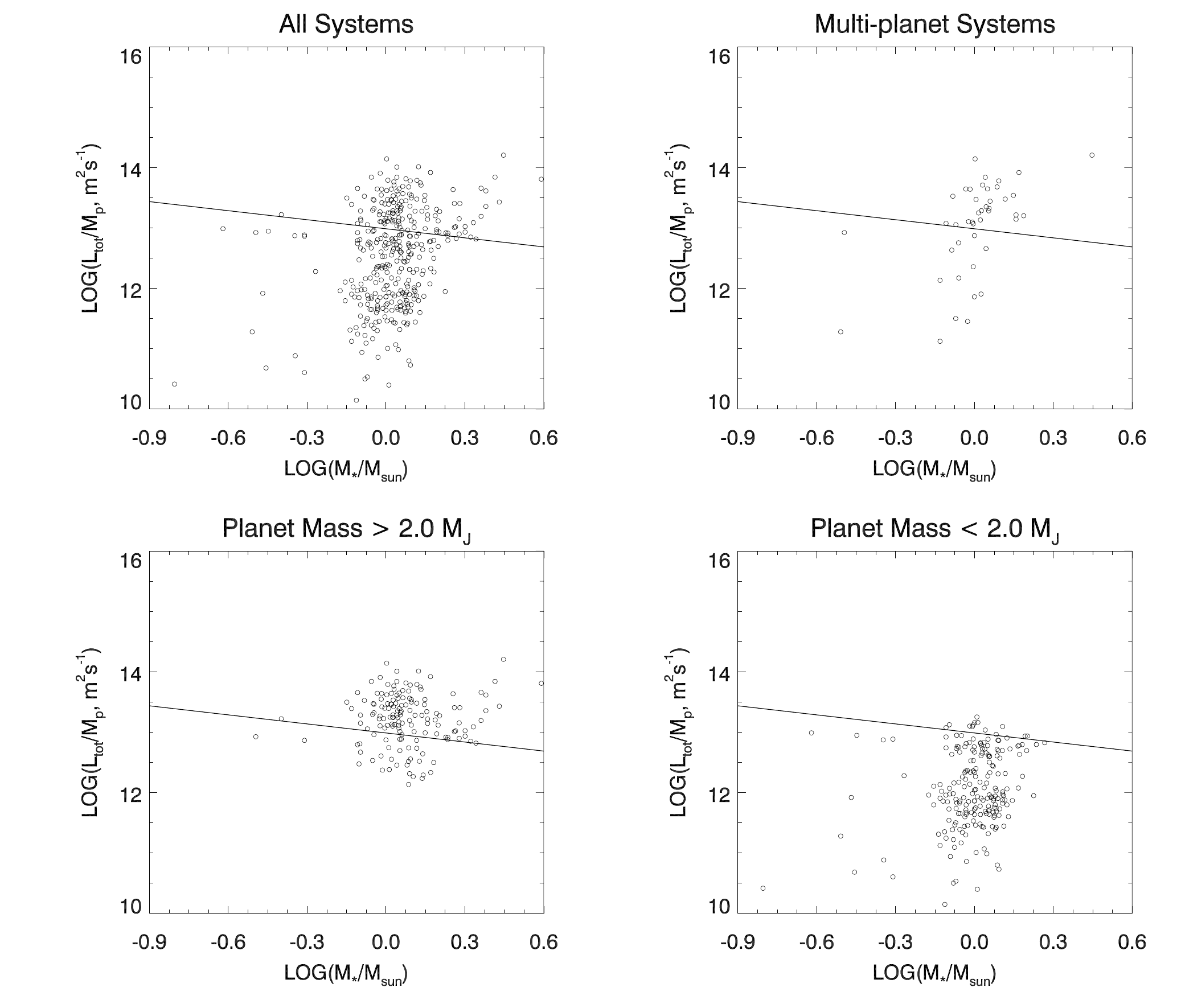}
\caption{Plot of the specific angular momentum, normalized to the stellar mass of each system, for all systems, multi-planet systems, systems with total planetary mass above 2.0 $M_J$, and systems with total planetary mass below 2.0 $M_J$.   The solid line represents a model of a Jupiter mass planet in a circular orbit at 5.2 $AU$ for comparison.}
\end{figure*}

\begin{figure*}
\epsscale{1.0}
\label{fig:specstarplan}
\plotone{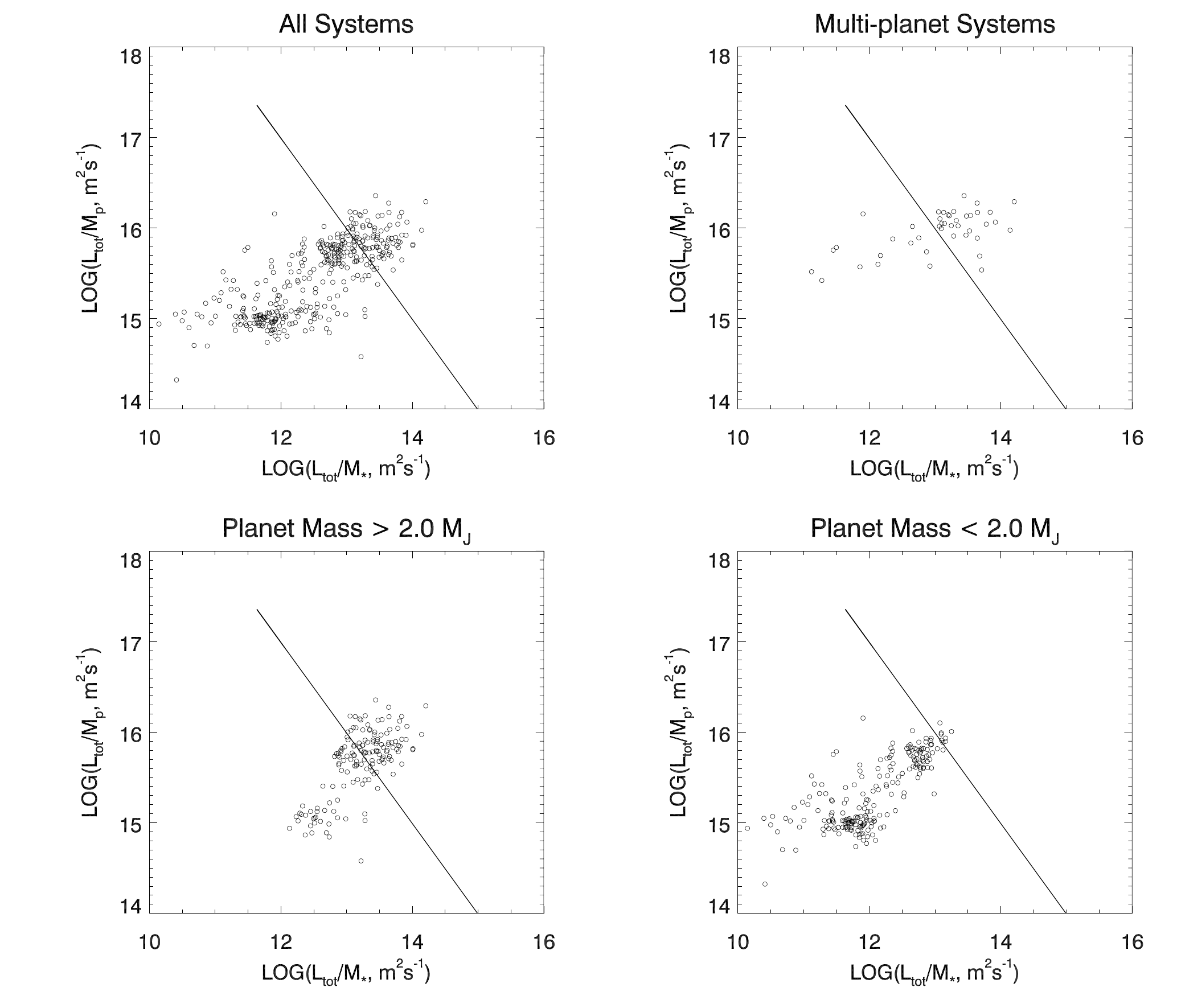}
\caption{Plot of the specific angular momentum, normalized to the stellar mass of each system, against the specific angular momentum, normalized to the system planet mass, for all systems, multi-planet systems, systems with total planetary mass above 2.0 $M_J$, and systems with total planetary mass below 2.0 $M_J$.   The solid line represents a model of a Jupiter mass planet in a circular orbit at 5.2 $AU$ for comparison.}
\end{figure*}

\begin{figure}
\epsscale{0.6}
\label{fig:hist}
\plotone{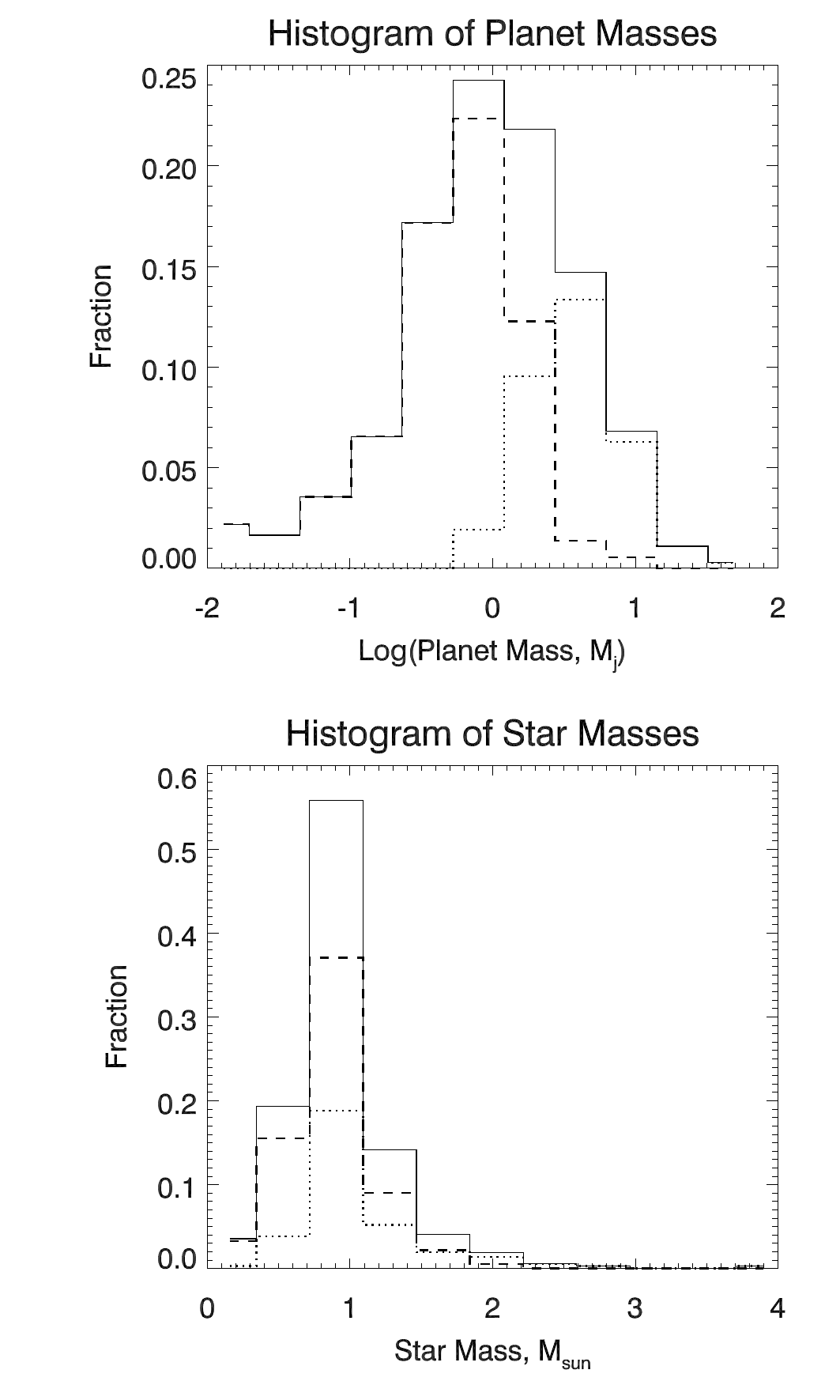}
\caption{Histogram showing the fraction of the total systems with measured parameters for the planets and stars in the sample.  The top panel is the planet mass (in units of $M_J$),  and the bottom panel is the stellar mass (in units of $M_{\odot}$).  The solid lines represent all the systems in the sample, the dashed lines are those systems with a specific angular momentum, $J_*$, less than $10^{13} \ m^{2} \ s^{-1}$, and the dotted lines are those systems with a specific angular momentum greater than $10^{13} \ m^{2} \ s^{-1}$.}
\end{figure}

\end{document}